\documentclass{iau} 
\usepackage{graphbox,graphicx}
\usepackage{natbib}
\usepackage{subfigure}
\usepackage[dvipsnames]{xcolor}
\usepackage{multirow} 
\usepackage{array,booktabs,arydshln,xcolor} 
\newcolumntype{P}[1]{>{\centering\arraybackslash}p{#1}} %
\newcommand{\argdf}{\texttt{ARGdf}}
\newcommand{\archain}{\texttt{AR-CHAIN }}

\title[S-stars Hosting Planets] 
{Interaction of Stars Hosting Planets \\ with Sgr A* Black hole}

\author[Nazanin Davari, Roberto Capuzzo Dolcetta \& Rainer Spurzem]   
{Nazanin Davari$^1$, Roberto Capuzzo-Dolcetta $^2$
	\and Rainer Spurzem$^3$}

\affiliation{$^1$Dept. of Physics, University of Rome La Sapienza, \\ P.le A. Moro 5, 00185, Roma, Italy, \\ email: {\tt nazanin.davari@uniroma1.it} \\[\affilskip]
	$^2$Dept. of Physics, University of Rome La Sapienza, \\ P.le A. Moro 5, 00185, Roma, Italy, \\ email: {\tt roberto.capuzzodolcetta@uniroma1.it} \\[\affilskip]	
	$^3$Astronomisches Rechen-Institut, Zentrum f{\"u}r Astronomie, \\ University of Heidelberg, M{\"o}nchhofstr. 12-14, 69120 Heidelberg, Germany \\email: {\tt spurzem@ari.uni-heidelberg.de}}

\pubyear{2019}
\volume{351}  
\jname{Star Clusters: From the Milky Way to the Early Universe}
\editors{A. Bragaglia, M.B. Davies, A. Sills \& E. Vesperini, eds.}
\begin{document}
	
	\maketitle
	
	\begin{abstract}
		We present some preliminary results of our ongoing project about planetary systems around S-stars in the vicinity of Sgr A* black hole. Since S-stars might have migrated in the Galactic Centre (GC) from elsewhere, they probably still keep their planetary systems throughout their voyage. In this work, we study the destiny of their putative planetary systems after close interaction with the central black hole of our galaxy.
	\end{abstract}
	
	\firstsection 
	\section{Introduction}
	
	The center of our galaxy is occupied with a dynamically relaxed dense cluster of stars, the so-called "S-stars", orbiting in the central gravitational potential which is believed to be created by a Supermassive black hole (SMBH) with a mass of $\sim$ 4.3 $\times {10^6} M_\odot$ at the distance $\sim$ 8.3 kpc from us \citetext{e. g., \citealt{gillessen2009}}. Recent spectroscopy analysis of the central arcsecond of the Galactic Center (GC) reveals the presence of a population of both early-type and late-type among the stars in the S-star cluster with the magnitude in the range $m_K=14-17$ (\citealt{habibi2017,habibi2019}). The age estimated for the star S2 is about 6.6 Myr and for the rest of the early-type stars is less than 15 Myr whereas for the late-type stars is $\sim$ 3 Gyr. The younger their ages the more mysterious seems their presence in their current location since strong tidal forces of the SMBH will inhibit star formation in that region (\citealt{morris1993}). For this reason, the migration model, due to binary disruption, is more likely to explain their presence in the vicinity of Sgr A* (\citealt{hills1988}). Therefore, it is usually assumed that the S-stars formed elsewhere and migrated to their current locations (\citealt{antonini2013}).\\
	The S-star cluster consists of roughly 40 stars. The brightest star of this ensemble is the star S2 which is measured to be a B-type main sequence star \citetext{e.g., \citealt{ghez2003}}. In general, 32 out of 40 stars of this group are young stars of spectral type of B0-B3V (\citealt{gillessen2017}) with the estimated mass of $8-14 M_{\odot}$ (\citealt{habibi2017}) while the other 8 stars are old (G, K, and M type) with an initial-mass range of $0.5 - 2 M_{\odot}$ (\citealt{habibi2019}).\\
	On the other hand, the existence of planets or planetary systems in the vicinity of the Sgr A* is still debated. \cite{trani2016} studied the tidal capture rate of single planets orbiting stars in the CW disk and in the S-star cluster. Moreover, many studies have done so far to investigate the dynamics and stability of planetary systems in star clusters \citetext{e.g., \citealt{spurzem2009,cai2017}}. Following the migration scenario, we suppose in this work that the S-stars still carry their planetary systems after travel to the GC and investigate the dynamics and stability of presumed planetary systems around these stars. 
	
	\begin{figure}
		\centering
		\includegraphics[align=t,scale=0.265]{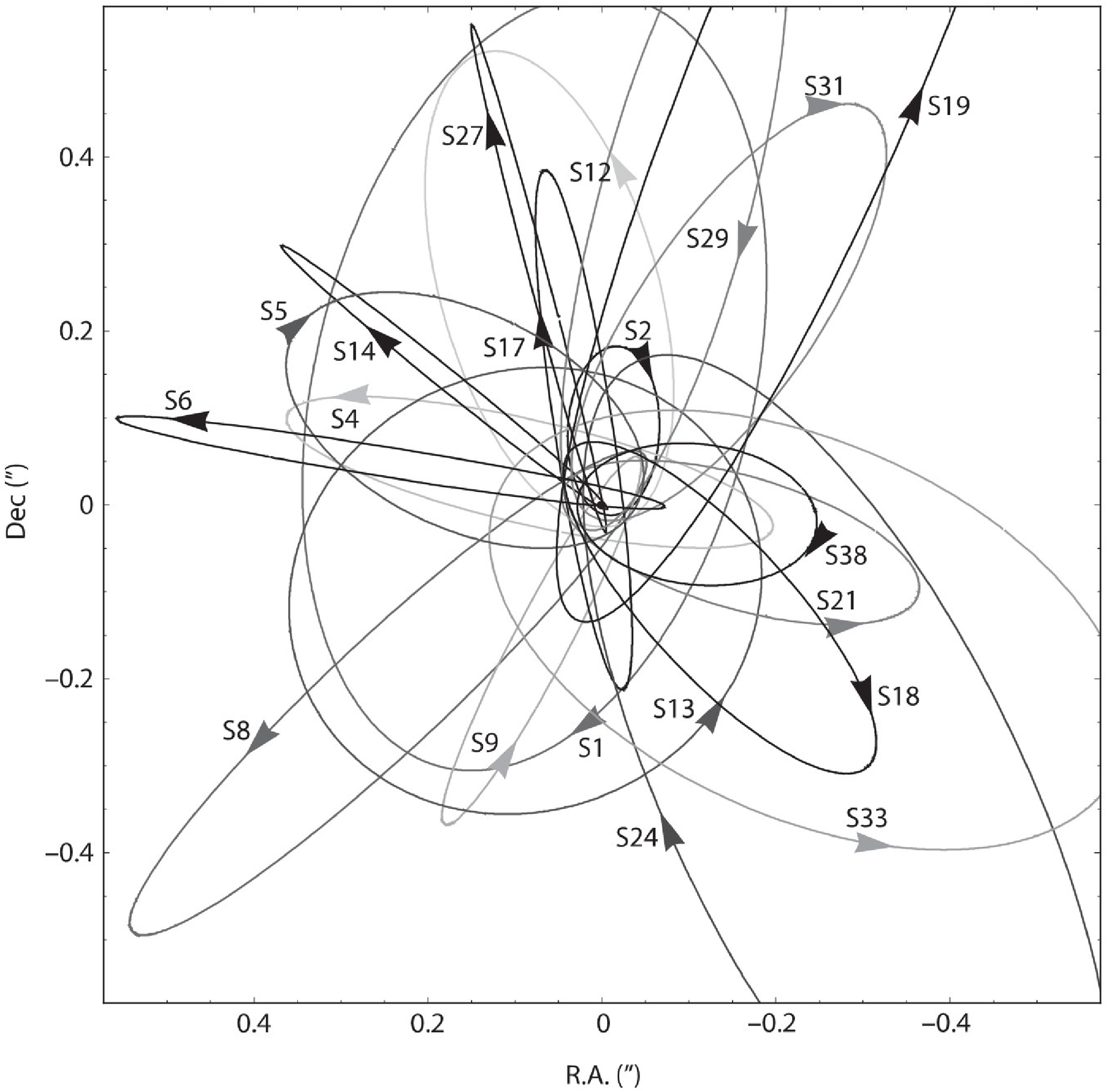}
		\includegraphics[align=t,scale=0.345,angle=90]{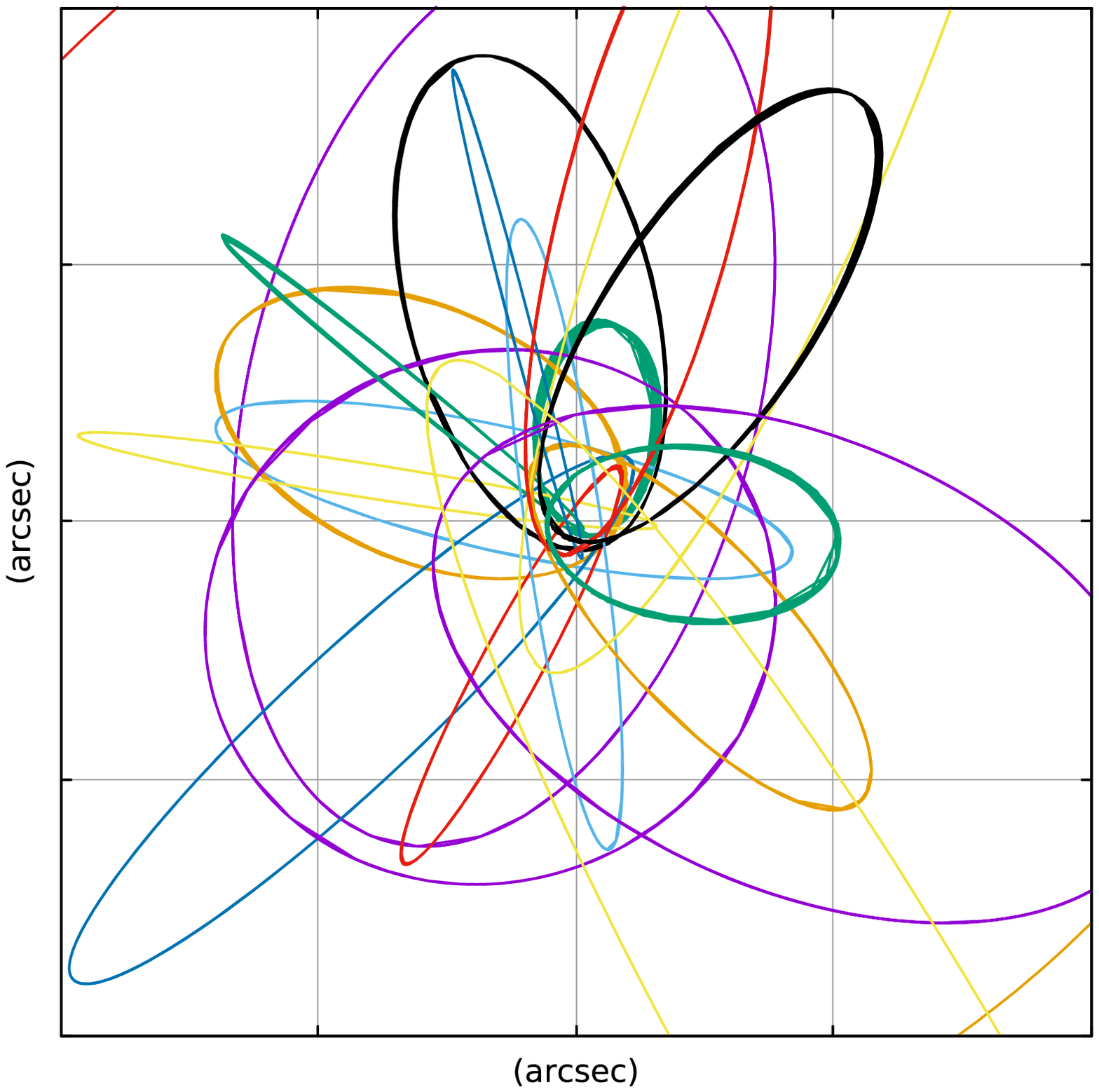}
		\caption{Comparison of trajectories of the S-stars projected onto the plane of the sky due to \cite{gillessen2009} (top) with their orbits due to our simulation including PN terms up to order 2.5 (bottom). The scale of the both figures is the same in arcsecond. \label{fig1}}
	\end{figure}

	\section{Method}
	
	We run simulations for the 40 innermost stars for which their orbital parameters are known, using the data of \cite{gillessen2009,gillessen2017} as initial conditions. We carry out our simulations by means of  \argdf code (\citealt{arcasedda2019}) which is a modified version of \archain (\citealt{mikkola1999,mikkola2008}), a fully regularized $N$-body code with post-Newtonian corrections up to order 2.5, enabled to use analytical external potentials and their dynamical friction. 
	The semi-major axis of the S-stars lies in the range of $\sim$ 0.005-0.05 pc and they approach very close to the Sgr A* where the relativistic terms could be important. For this reason, we made our simulations taking into account the post-Newtonian corrections of equation of motion up to order 2.5.  \\
	The mass of Sgr A* is set to $\sim$ 4.3 $\times {10^6} M_\odot$ and a distance of 8.33 kpc to Sgr A* is assumed (\citealt{gillessen2009,gillessen2017}), i.e., 1" corresponds to 40.4 mpc (milliparsec) (\citealt{merritt2013}). Both population of young (B-type) or old (G, K, and M type) giant stars (post main-sequence) are observed in this region \citetext{e.g., \citealt{ghez2003}}. The stars in \cite{habibi2017} belong to the early-type group
	and in \cite{habibi2019} belong to the late-type group. \cite{habibi2017} derived specifically the masses of 8 stars in early-type group. We use their evaluation to estimate the mass of other early-type stars which have a similar magnitude ($m_k$) to the stars which are studied in \cite{habibi2017} and for the rest of early-type S-stars (fainter ones) we assume the mass range of $8-14 M_\odot$. Furthermore, \cite{habibi2019} measured an initial-mass range of $0.5-2 M_\odot$ for the older stars ($\sim$ few Gyr old). We also implement the mass range of $0.5 - 2 M_\odot$ for late-type stars in our simulations.\\
	We assign planetary systems to each of the stars in S-star cluster similar to our Solar planetary system in mass, eccentricity and semi-major axis. For late-type stars, every planetary system is composed of 7 planets; Mercury, Venus, Earth, Mars, Jupiter, Saturn and Uranus. For the early-type stars, every planetary system consists of 3 planets ( Jupiter, Saturn and Uranus). Since it is less likely that massive early-type stars could harbor planets so close, for these stars we exclude the first 4 planets.\\
	We change the inclination of the planetary system orbits respect to the parent star orbital plane in the range $0^\circ-180^\circ$ with increments of $10^\circ$. Since the star S85 has the longest orbital period of $\sim$ 3580 yr, we stop the simulations at the time 4000 yr to follow at least one of its periods. The average computation time for each run is about 48 hours.
	Table \ref{tab1} shows the initial parameters of our runs. Fig. \ref{fig1} gives snapshot comparison of our computed orbits of the S-stars (without planets) with the observed trajectory of S-stars in \cite{gillessen2009}.

	\begin{table}
		\begin{center}
			\caption{Assumed parameters for our simulations.}
			\label{tab1}
			{\scriptsize
				\begin{tabular}{|l|c|c|c|c|}\hline 
					{\bf Spectral} & {\bf Mass } & {\bf Number } & {\bf inclination} \\ 
					{\bf Type} & {\bf  Range} & {\bf of} &  {\bf (degree)} \\
					&({\bf  $M_{\odot}$})& {\bf Planets} &  \\ \hline
					early-type & $8-14 $ & 3 &  $0-180$ \\ \hline
					late-type & $ 0.5-2$ & 7 &  $0-180$  \\ \hline
				\end{tabular}
			}
		\end{center}
		\vspace{1mm}
	\end{table}

	\begin{figure}[t]
		\begin{center}
			\includegraphics[width=3.6in]{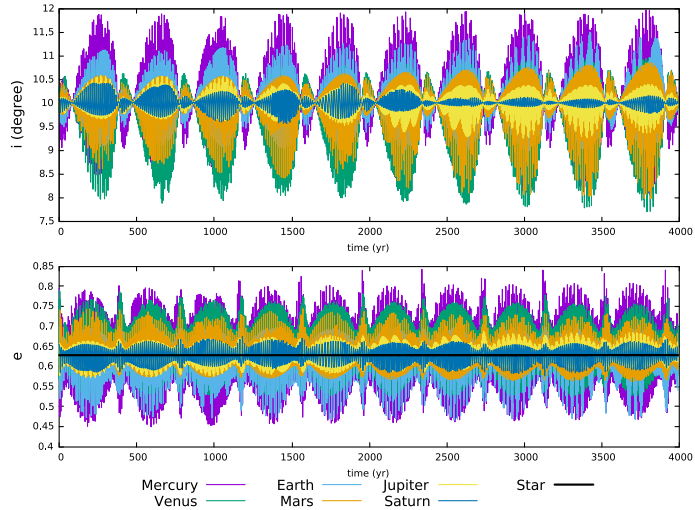} 
			\caption{The stability of orbital elements versus time for "S89" planetary system. The black line shows the eccentricity of the host star "S89". }
			\label{fig3}
		\end{center}
	\end{figure}

	\section{Preliminary Results}
	
	Our aim is to study dynamical evolution and stability of hypothetical planetary systems around the S-stars in the vicinity of the Sgr A* in the GC. Some preliminary results of our simulations are shown in Table. \ref{tab2}. For this set, the masses of early-type stars which are neither reported in \cite{habibi2017} nor comparable with their estimated masses in magnitude, are set to $12  M_{\odot}$ and the masses of late-type stars are set to $1.5 M_{\odot}$.\\ Our simulations shows that the innermost planets i.e, Mercury, Venus, Earth and Mars are more likely to stay bound to their host stars and almost all the Mercury-type planets remain bound to their late-type parent stars until the end of simulation (see Table. \ref{tab2}). Moreover, some of the early-type stars escape the GC and their planetary systems are swapped by a nearby star.\\
	Fig. \ref{fig3} illustrates the evolution of orbital parameters of the star "S89" planetary system. The system initially has 7 planets which are in co-planar orbits but with inclination $10^\circ$ to their host stars. six out of seven planets remain bound to the star "S89" until the end of simulation.  
	
	\begin{table}
		\begin{center}
			\caption{Some preliminary results of our simulations for the inclinations in the range of $0^{\circ}-90^{\circ}$ (the rates are in $\%$). The first column shows spectral type of the host S-stars (SP) which are indicated as "l" and "e" for "early" and "late"-type S-stars. Note that "early-type" stars do not host the first four planets.} 
			\label{tab2}
			\begin{tabular}{|c|c|c|c|c|c|c|c|}
				\hline
				\multicolumn{8}{|c|}{{\bf Bound Planets to the S-stars}}\\ \hline
				SP & Mercury & Venus & Earth & Mars & Jupiter & Saturn & Uranus \\
				\hline
			l & 100.00 & 72.50 & 72.50 & 61.85 & 48.75 & 36.25 & 16.25  \\ 
			e & - & - & - & - & 75.62 & 47.50 & 35.31 \\ \hline
			\end{tabular}
		\end{center}
		\vspace{1mm}
	\end{table}

\end{document}